%% file: main.tex
\documentclass[fleqn,10pt]{wlscirep}
\usepackage[utf8]{inputenc}
\usepackage[T1]{fontenc}
\usepackage{tikz-cd}
\usepackage{wrapfig}
\usepackage[most]{tcolorbox}
\usepackage{float}
\tcbset{
  colback=gray!5,
  colframe=black!60,
  fonttitle=\bfseries,
  coltitle=black,
  boxrule=0.8pt,
  arc=2pt,
  outer arc=2pt,
  left=6pt,
  right=6pt,
  top=6pt,
  bottom=6pt,
}

\usepackage[most]{tcolorbox}
\usepackage{xparse}
\newtcolorbox{dialogbox}[2][blue]{%
  arc=0pt,
  left=2mm, right=2mm,
  top=1mm, bottom=1mm,
  boxrule=0.4pt,
  colback=white,
  colframe=#1,               
  breakable,
  title=#2,
  fonttitle=\bfseries,
  toprule=0.8pt              
}
\usepackage{adjustbox}
\usepackage{enumitem}

\usepackage{listings}
\tcbuselibrary{skins,breakable}
\lstdefinestyle{promptstyle}{
  basicstyle=\ttfamily\small,
  breaklines=true,            
  breakatwhitespace=false,    
  columns=fullflexible,       
  keepspaces=true,            
  showstringspaces=false,
  inputencoding=utf8,         
}
\newtcolorbox{promptbox}[2][]{%
  enhanced,
  breakable,
  colback=gray!5,
  colframe=blue!60!black,   
  colbacktitle=blue!70,     
  fonttitle=\bfseries,
  coltitle=white,           
  boxrule=0.8pt,
  arc=2pt,
  outer arc=2pt,
  left=6pt, right=6pt, top=6pt, bottom=6pt,
  title=#2,
  #1
}

\usepackage{longtable}
\usepackage{graphicx}
\usepackage{makecell}

\title{Multi-Agentic Approach for History Matching of Oil Reservoirs}

\author[1]{Linar Samigullin}
\author[1, *]{Sergei Shumilin}
\author[1,2]{Evgeny Burnaev}
\affil[1]{Skoltech, AI Center, Moscow, Russia}
\affil[2]{AIRI, Moscow, Russia}

\affil[*]{s.shumilin@skoltech.ru}

\begin{abstract}
History matching is a central inverse problem in reservoir engineering, where uncertain reservoir parameters must be calibrated against observations. Although automated history matching can reduce manual effort, practical deployment remains difficult because engineers must still configure heterogeneous workflows involving parameter selection, physically admissible bounds, optimizer choice, hyperparameter tuning, simulator execution, and diagnostic reporting. We propose \textbf{PetroGraph}, a multi-agent framework for intelligent reservoir history matching that decomposes this workflow into specialized agents for model review, experimental planning, parameterization, optimization, simulation, and summarization. The system combines large language model agents with domain-specific tools, retrieval-augmented access to simulator documentation, validation of modified ECLIPSE input decks, human-in-the-loop checkpoints, and an OPM Flow-based simulation backend. This design enables users to initiate and steer history matching through natural language while preserving explicit control over selected parameters and optimization settings. We evaluate \textbf{PetroGraph} on three reservoir models of increasing complexity: the synthetic SPE1 model, the faulted SPE9 benchmark, and the real-field Norne model. Using weighted normalized root mean square error as the objective, \textbf{PetroGraph} reduces the mismatch by 95\% on SPE1, 69\% on SPE9, and 13\% on Norne. These results demonstrate that multi-agent orchestration can automate key decisions in history matching, lower the expertise barrier for operating complex simulation workflows, and provide a flexible foundation for extensible, domain-aware reservoir model adaptation.

\end{abstract}
\begin{document}

\flushbottom
\maketitle
%
%
\thispagestyle{empty}

\section*{Introduction}

\begin{wrapfigure}{r}{0.55\textwidth} 
    \centering
    \includegraphics[height=0.3\textwidth]{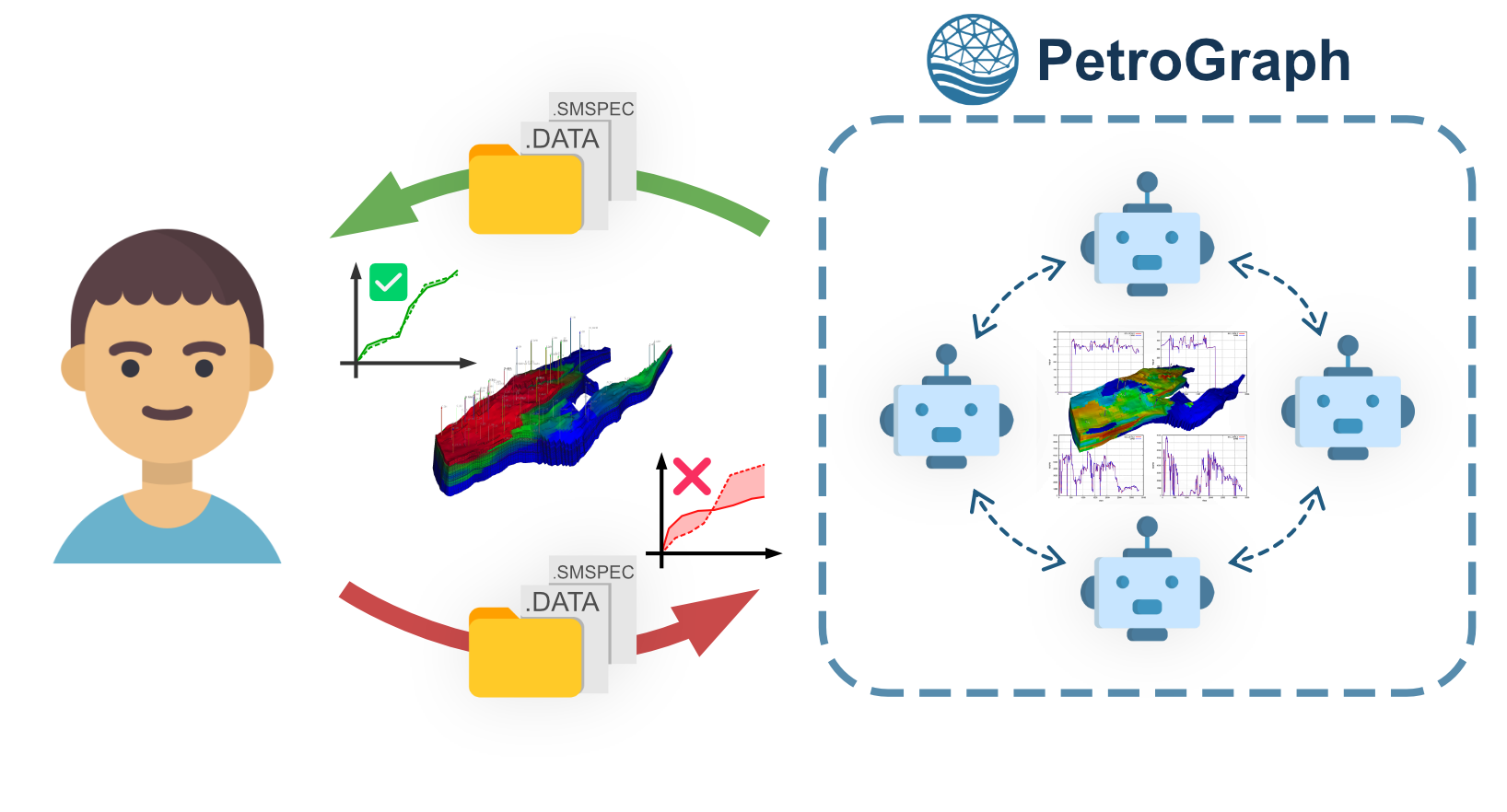} 
   \caption{Conceptual overview of \textbf{PetroGraph}. A user interacts with a reservoir model through natural language; \textbf{PetroGraph} translates the request into a coordinated multi-agent workflow that reviews input data, selects parameters and bounds, runs optimization and simulation, and returns validated model updates and diagnostic feedback.}
\end{wrapfigure}

An important applied domain in which methodological heterogeneity poses a significant challenge is oil reservoir model management. Oil companies typically employ large teams of specialists responsible for updating numerical reservoir models as new field data become available. This process is labor-intensive and strongly dependent on the expertise and experience of reservoir and petroleum engineers.

The task of calibrating a reservoir model against observed field data, such as oil production rate, water production rate, bottom-hole pressure, reservoir pressure, well-log measurements, and other observations, is referred to as reservoir model history matching. History matching consists of iteratively modifying model parameters and evaluating the discrepancy between simulation outputs and observed field data. This process can be automated. Automatic history matching belongs to a class of inverse problems in which the objective is to identify unknown parameters of a mathematical model that allow it to reproduce observed data as accurately as possible. Such problems are typically formulated as optimization problems \cite{Samoil2024} and involve heterogeneous components, including optimization algorithms, parameterization techniques, and numerical solvers for reservoir flow simulation.

The central challenge addressed in this study lies in the diversity of available methods and their software implementations. Multiple optimization techniques are applicable to reservoir inverse problems, and each often requires a number of hyperparameters to be selected or tuned for a specific scenario. This creates a problem of methodological multiplicity, which is the core focus of the present research. Modern artificial intelligence tools, particularly multi-agent systems, provide a way to address this challenge through an LLM-based orchestrator capable of selecting appropriate methods and configuring their parameters in an informed manner.

The application of a multi-agent approach to automatic reservoir history matching is particularly natural because the main components of the history-matching workflow can be decomposed into specialized tasks. These tasks can be represented as independent agents, each with a clearly defined objective. For example, an optimization agent can select a suitable optimization method for a given task and configure its hyperparameters. Moreover, the system can interact with the user through natural language, thereby reducing the expertise required to operate complex history-matching workflows. A functional multi-agent system for oil reservoir history matching can also be extended to related application scenarios, such as gas reservoir history matching or calibration of geological models for mineral deposits, making the approach suitable for multi-domain engineering applications.

In practice, engineers face the difficult problem of selecting an optimal computational pipeline for a given history-matching task. Even experienced engineers may find it challenging to account for the full range of modern methods, software tools, and configuration choices. We aim to address this problem by developing a system that simplifies automatic history matching through a multi-agent methodology.
We present \textbf{PetroGraph}, a framework for intelligent automatic history matching. \textbf{PetroGraph} provides a user-friendly interface that enables engineers without a computer science background to configure, execute, and analyze history-matching workflows more easily.

\section*{Related Work}

\paragraph{History matching (HM) of oil reservoirs.} 
History matching is an inverse problem that aims to identify a set of reservoir parameters that best fits real-world field data.
These parameters are usually divided into two categories: static parameters, such as permeability and porosity, and dynamic or flow-related parameters, such as relative permeability.
In the optimization-based formulation of history matching, the reservoir simulator is commonly treated as a black-box function whose outputs are compared with observed data.
Usually, an engineer selects the parameters to be optimized based on the analysis of the model.
HM has been extensively studied since the 1990s, and today the primary research focus is concentrated on automated history matching \cite{Oliver2011ReservoirHistoryMatching}.
In practice, industrial oil reservoir models may contain tens of millions of cells, which makes hydrodynamic simulation computationally expensive.
This motivates reducing the number of simulator runs required during optimization.
One of the most advanced approaches to sample-efficient optimization is Bayesian optimization (BO), which builds a surrogate model of the black-box objective function using a Gaussian process model and efficiently queries this model to select the next parameter set for evaluation \cite{CHAI2021108204, Samoil2024}.

\paragraph{Multi-agent systems (MASs).} 
MASs are currently demonstrating rapid growth in both the number of publications and the range of practical applications. 
The combination of a system prompt, an LLM, and a set of tools has become a de facto practical implementation of an agent.
The main motivation comes from the idea that a set of diverse cooperating agents can be more efficient than a single actor.
The key properties of MASs---the autonomy of agents, their flexibility, and their ability to adapt to environmental changes---allow them to solve complex distributed problems by decomposing them into subtasks executed by individual agents \cite{naji2020mas}.

In this study, the definition of an agent is adopted from the classical work \cite{wooldridge1995agents}.
An LLM-based agent can dynamically form its actions using generalized knowledge acquired through pre-training on large datasets.
In 2020--2024, the first works demonstrating a new approach appeared: instead of following a rigidly defined behavioral program, the agent is supplied with a prompt describing its goal and context, while control over its actions is delegated to a language model via a set of tools \cite{jimenezromero2025llmmas}.

\paragraph{MASs in engineering.}
Current research in this area demonstrates the effectiveness of MASs in solving engineering problems through the cooperation of specialized agents. 
As the authors of \cite{chen2024survey} note, this architecture provides a more accurate representation of the real world compared with single-agent systems, since many applied problems naturally involve the interaction of multiple participants making decisions simultaneously. 
In \cite{elrefaie2025car}, the authors present a MAS for optimizing automobile design. A set of agents iteratively modifies the shape of the car body and then calculates the aerodynamic characteristics of the resulting shape. 
This system is a clear example of a hybrid multi-agent system, as it combines agents based on deep learning with agents that perform numerical simulations.

In the context of BIM design automation, it has been shown that LLM agents effectively interpret unstructured design requirements and transform them into API calls for authoring systems, while rule-based validators and geometric modules provide formal verification and correction of models \cite{du2024text2bim}.
In automotive engineering, agent-based LLM systems demonstrate the ability to accelerate HARA analysis, requirements generation, and test planning. This is achieved by combining LLM-based semantic inference with rigorous validation of results obtained by traditional methods. This hybrid approach reduces the labor intensity of documentation preparation and improves the reproducibility of safety processes \cite{shi2024aegis}.

In manufacturing systems, LLM-enabled MASs are used for process coordination: the LLM agent distributes tasks, generates control instructions, and adjusts the process flow when requirements change. 
Specialized agents perform calculation procedures and optimization steps, which increase system flexibility and reduce equipment changeover time \cite{lim2024manufacturing}.

In the energy sector, frameworks are being developed in which LLM agents perform semantic analysis of events and generate corrective scenarios, while numerical modules simulate and verify these scenarios before execution. This approach shows potential for faster detection of network disturbances \cite{zhang2025gridagent}.

For human--robot collaboration (HRC) tasks, LLM-oriented approaches enable operators to specify requirements in natural language, and the agent system transforms them into a sequence of controlled steps and robot programs, while digital-twin and verification agents ensure safe execution and compliance with technological constraints \cite{gkournelos2024hrc}.

The work \cite{jbde.2025.0018} addresses the problem of coordinated decision-making at various project stages in the construction industry. The proposed MAS, in which heterogeneous agents representing different disciplines, such as design and finance, engage in a structured dialogue to identify and resolve conflicts, involves exchanging evidence and developing a collective decision. This system demonstrated its effectiveness in four scenarios, significantly improving cost-forecasting accuracy and reducing the number of design flaws. Other applications of MASs for solving design problems in the construction industry are discussed in \cite{Alada2019MultiAS}.

The work \cite{mushtaq2025harnessing} addresses the problem of integrated support for engineering work, taking into account technical, ethical, and organizational requirements. The authors propose a MAS architecture in which each LLM-based agent plays the role of a subject-matter expert, while a coordinating mechanism ensures collaborative decision-making. The approach has been tested on a number of real-world project proposals, where it has been shown to provide more comprehensive and balanced recommendations than a single agent, improving the quality and completeness of engineering solutions.

The study \cite{yan2025build} addresses the low efficiency of conventional LLMs when solving engineering problems that require both domain knowledge and multi-step computations. The authors propose a hybrid MAS architecture in which one agent uses RAG to improve the accuracy of theoretical answers, while another performs mathematical modeling under the control of a central coordinator. This approach provides more robust and interpretable solutions to engineering and biochemical problems, significantly outperforming basic LLMs in terms of stability and completeness of results.

The topic of context-aware MASs is addressed in the study \cite{du2024comprehensive}. This paper presents a detailed overview of context-aware MASs, describes key approaches to their design, and presents a wide range of engineering applications of such systems, including smart cities, supply chains, and other applications where agents must adapt to a dynamic environment. Particular attention is also paid to reliability and security in such heterogeneous environments.

\paragraph{MASs interacting with physical solvers.}
Several recent studies have explored the development of MASs designed to interface with physical solvers. For instance, the work presented in \cite{yue2025foamagent20endtoendcomposable} introduces a multi-agent framework for computational fluid dynamics (CFD) that generates an end-to-end OpenFOAM workflow from a single natural-language prompt. Evaluated on a benchmark dataset covering a wide variety of CFD simulation types, the system achieves a success rate of 88.2\%. The authors attribute this significant improvement over existing frameworks---such as MetaOpenFOAM \cite{chen2024metaopenfoamllmbasedmultiagentframework}, which achieves only a 55.5\% success rate on the same benchmark---to the implementation of a Hierarchical Multi-Index RAG system. This system is constructed from OpenFOAM tutorial cases, allowing for more precise and contextually relevant information retrieval.

Frameworks such as AutoFEA \cite{Hou_Johnson_Makhija_Chen_Ye_2025} and MooseAgent \cite{zhang2025mooseagentllmbasedmultiagent} have demonstrated the ability to translate natural-language descriptions of structural engineering problems into executable input files for finite element analysis (FEA) solvers, such as CalculiX and Abaqus. A key innovation in AutoFEA is its use of a GCN--Transformer link-prediction model for retrieval, which accounts for the sequential dependencies between steps in a simulation workflow.

\section*{Methodology}

\subsection*{Multi Agentic Approach}

\textbf{PetroGraph} is a multi-agent system designed to accelerate and lower the entry barrier for the critical industry process of history matching in oil and gas field development.

The process begins with loading the reservoir data in the ECLIPSE~100 format, which consists of a primary .DATA file and its associated files, along with the corresponding simulation results stored in .SMSPEC files. These files contain both the initial simulation outputs and the actual field data against which the history matching will be performed.

\begin{figure}[ht]
    \centering
    \includegraphics[width=\linewidth]{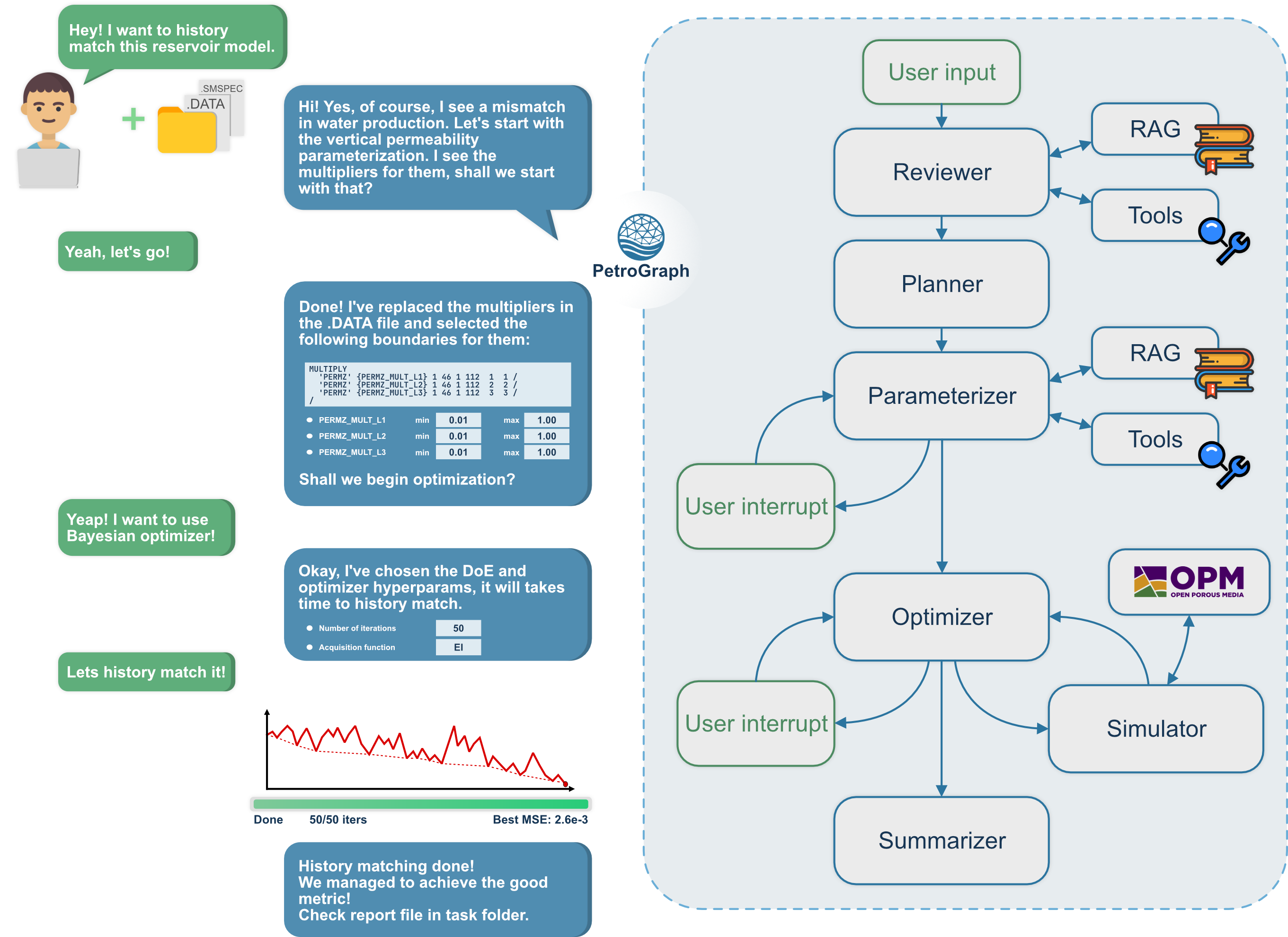}
    \caption{\textbf{PetroGraph} system architecture illustrating end-to-end example workflow. The system features six agents with human-in-the-loop: 1. \textbf{Reviewer Agent}: examines the reservoir model and creates a textual representation, 2. \textbf{Planner Agent}: analyzes data mismatch and proposes a DoE for subsequent agents, 3. \textbf{Parameterizer Agent}: finds and sets parameters for history matching and changes keywords content,  2. \textbf{Optimizer Agent}: chooses DoE and hyperparameters for optimizer, 3. \textbf{Simulator Agent}: evaluates simulations and calculates metric, 4. \textbf{Summarizer Agent}: makes summary of reservoir history matching results.}
    \label{fig:chat_and_schema}
\end{figure}

The system employs a hybrid interface that combines primary interaction through a chat with manual input fields for parameter editing and selection (see Fig.~\ref{fig:chat_and_schema}). This approach preserves the familiar chat-based interaction model common to agentic systems while alleviating the need for the user to repeatedly ask the system to change specific parameters. Instead, it enables users to make adjustments directly.

The multi-agent framework consists of several agents linked in an execution chain, which can be represented as a graph (see Fig.~\ref{fig:chat_and_schema}). These agents are the Reviewer Agent, Planner Agent, Parameterizer Agent, Optimizer Agent, Simulator Agent, and Summarizer Agent. All agents share information through a common state. Detailed descriptions of each agent are provided in the corresponding subsection. The framework is based on the open-source LangGraph framework. A detailed graph architecture is provided in Appendix~\ref{app:graph}.

\subsubsection*{Agents}

The \textbf{Reviewer Agent} is an LLM-based agent. It is the first to respond: it investigates the provided reservoir model using its tools and then creates a concise textual summary of the data for subsequent agents. This summary includes the model type (blackoil, dissolved gas, etc.), model dimensions, and the presence of multipliers, faults, and other features relevant to history matching.

The \textbf{Planner Agent} is an LLM-based agent. It receives global field metrics -- provided separately for production, injection, fluid type, and bottom-hole pressure -- together with the concise description from the Reviewer Agent. Its task is the Design of Experiment (DoE): it recommends which parameters to select for history matching and defines the optimisation strategy while explicitly accounting for computational cost. For lightweight models the number of iterations and parameters can be generous; for heavier models a conservative approach is warranted in selecting both the number of parameters and iterations.

The \textbf{Parameterizer Agent} is an LLM-based agent. It receives the brief reservoir description and the plan from the Planner Agent. It begins by conducting an in-depth investigation of the input-deck structure, then builds a representation of the model by identifying parameters that can be varied and defining their physically and context based ranges. Template files containing parameter placeholders are created for future value insertion. Throughout this process the agent validates the input deck by performing dry runs with the minimum and maximum substituted values; any error (e.g., a non-monotonic relative permeability curve or an arithmetic expression inside a keyword that does not support it) is relayed back, preventing mistakes that the probabilistic model might otherwise make. The agent also has access to a RAG system based on the description of keywords and sections with examples of the ECLIPSE~100 .DATA file structure from the OPM simulator manual~\cite{flowmanual}. Once an error-free parameterization is obtained, a Human-in-the-Loop (HITL) checkpoint allows the user to adjust parameters and their bounds on the fly.

The \textbf{Optimizer Agent} is also an LLM-based agent. It receives the problem dimensionality and the DoE from the Planner Agent, then selects the DoE parameters -- including the number of iterations and the initial sampling strategy -- and chooses the optimizer type together with its hyperparameters. After initialisation, another HITL checkpoint occurs, at which the user may modify specific parameter values.

The \textbf{Simulator Agent} is the only agent that does not utilise an LLM. During the history-matching loop it operates in an ask-parameters/tell-metric format: it prepares the data for model execution, substitutes the parameter values provided by the optimizer, runs the OPM simulator~\cite{opm}, and calculates the resulting metrics.

To evaluate the quality of the history match, we employ the weighted mean normalized root mean square error (wNRMSE). For each well and for production and injection metrics, the weight is defined as the fractional contribution of the corresponding quantity to the total historical production or injection. For the bottom-hole pressure metric, the weight is defined as the reciprocal of the number of wells. The overall objective function is computed as the sum of contributions across all variables with nonzero historical data, including WBHP, WOPR, WWPR, WGPR, WOIR, WWIR and WGIR, with wNRMSE formulated as shown in Eq. \ref{eq:wNRMSE}.

\begin{equation}
\text{wNRMSE} = \text{weight} \cdot \text{NRMSE} = \text{weight} \cdot \frac{\text{RMSE}}{\bar{y}_{\text{hist}}} = \text{weight} \cdot \frac{ \sqrt{ \frac{1}{N} \sum_{i=1}^{N} {(y_{\text{sim}} - y_{\text{hist}})^2} } }{\bar{y}_{\text{hist}}}
\label{eq:wNRMSE}
\end{equation}

The \textbf{Summarizer Agent} is an LLM-based agent. Its primary task is to summarize the entire workflow that has been performed, to generate visualizations of selected plots, and to produce a summary report. If the objectives established by the Planner Agent are not met, the report includes recommendations -- for example attributing the outcome to a limited number of iterations (when that number was forcibly reduced via HITL) or to the absence of critical parameters from the history-matching process.

The system prompts for all LLM-based agents are provided in Appendix~\ref{app:prompts}.

\subsubsection*{Tools and RAG}

The \textbf{Reviewer} and \textbf{Parameterizer Agents} relies on a set of tools designed to explore the ECLIPSE input deck. The hierarchical structure of the deck allows it to be divided into sections. Each section contains keywords with associated content. Therefore, the toolset must include functions for retrieving the list of sections, obtaining the list of keywords within a given section, and fetching the content of a specific keyword. A RAG component is also integrated at this stage. It provides pre-processed descriptions of sections and keywords extracted from the OPM Flow Manual. This integration enables interaction with the LLM despite the model lacking fine-tuning on petroleum simulation data, while also reducing the probability of hallucinations.

The \textbf{Parameterizer Agent} also has access to tools for modifying keyword content. These tools allow for the insertion of placeholders. They also enable the addition and removal of parameters along with their variation ranges.

These tools are also utilized by the \textbf{Planner} and \textbf{Optimizer Agents} to generate structural output. For the Planner Agent, this output defines the division of the Design of Experiments (DoE) workflow between the Parameterizer and Optimizer agents. For the Optimizer Agent, this output populates the optimizer hyperparameters and specifies the number of iterations.

\subsection*{Data Model}

History matching experiments were conducted using models from the OPM data repository~\cite{opm_data_repo}. Three models were selected: pure synthetic SPE1, synthetic SPE9 with varying permeability, and real Norne field model.

\subsubsection*{SPE1}

\begin{figure}[ht]
    \centering
    \includegraphics[width=0.4\linewidth]{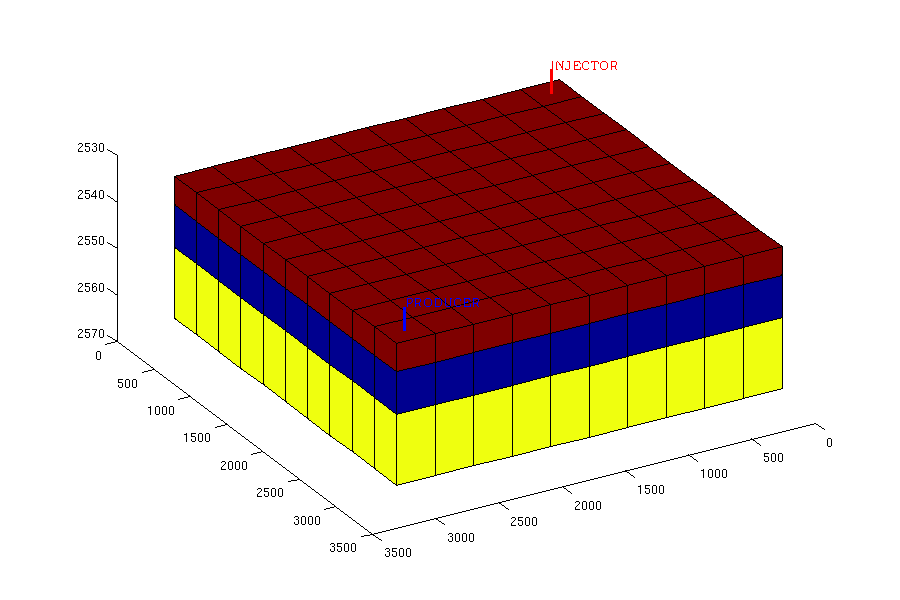}
    \caption{SPE1 simple synthetic reservoir model with two wells: one producer and one injector.}
    \label{fig:spe1_view}
\end{figure}

The SPE1 benchmark consists of a three-dimensional Cartesian grid with 300 active cells, distributed as $10 \times 10 \times 3$ in the X, Y, and Z directions, as illustrated in Fig.\ref{fig:spe1_view}. A single gas injection well is placed at grid block $(x=1,\,y=1,\,z=1)$, and a single producer is located at $(x=10,\,y=10,\,z=1)$.

The operational limits follow the definitions given by Odeh \cite{spe1_odeh} and the OPM data repository \cite{opm_data_repo}. The gas injector is constrained by a maximum surface gas rate of 100~MMscf/day and a maximum bottom-hole pressure (BHP) of 9014~PSIA. The production well operates under a maximum oil rate of 20,000~STBO/day and must maintain a minimum flowing BHP of 1000~PSIA.

\subsubsection*{SPE9}

\begin{figure}[ht]
    \centering
    \includegraphics[width=0.4\linewidth]{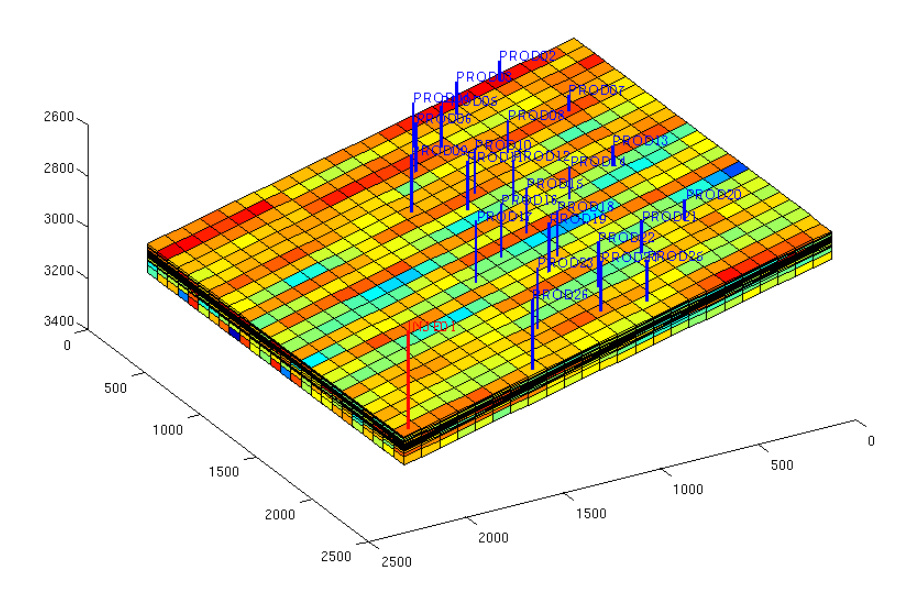}
    \caption{SPE9 complex synthetic reservoir model with 26 wells: 25 producers and one injector.}
    \label{fig:spe9_view}
\end{figure}

The SPE9 model, introduced by Killough \cite{spe9_killough}, is a three-dimensional reservoir simulation case containing 9,000 cells, arranged as $24 \times 25 \times 15$ in the X, Y, and Z directions. The field comprises 25 production wells and a single water injection well; the layout is displayed in Fig.\ref{fig:spe9_view}.

Well controls are based on the specifications of Killough \cite{spe9_killough} and the OPM data repository \cite{opm_data_repo}. The water injector is limited to a maximum water injection rate of 5,000~STBW/day and a maximum BHP of 4,000~PSIA, referenced to a depth of 9,110~ft. Each producer is initially subject to a maximum oil rate of 1,500~STBO/day, with a minimum flowing BHP of 1,000~PSIA at the same reference depth of 9,110~ft.

\subsubsection*{Norne}

The Norne Field is an oil-and-gas accumulation located in the Norwegian sector of the North Sea \cite{chen_norne_hm}. The stratigraphic section comprises five formations, listed from top to base: Garn, Not, Ile, Tofte, and Tilje. The simulation model captures four main fault blocks that are in partial hydraulic communication, as well as numerous internal faults whose connectivity is uncertain within each block. The field has been developed with 22 production wells and 9 injection wells, employing water-alternating-gas (WAG) injection as the primary recovery strategy. 

\begin{figure}[ht]
    \centering
    \includegraphics[width=0.4\linewidth]{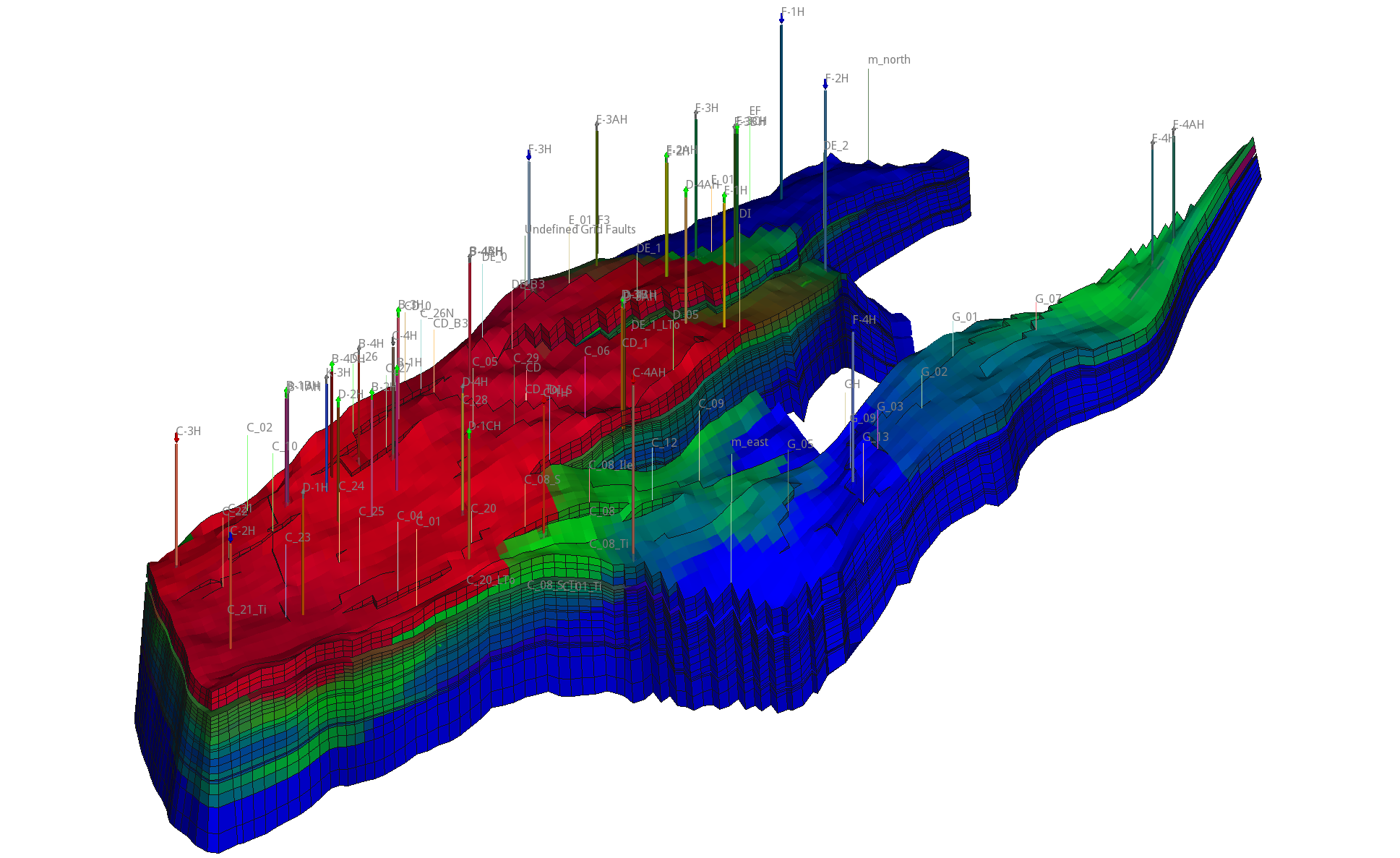}
    \caption{Norne field reservoir model with 31 wells: 22 producers and 9 injectors.}
    \label{fig:norne_view}
\end{figure}

The available dataset from the OPM repository \cite{opm_data_repo} is a manually history-matched model. Its grid dimensions are $46 \times 112 \times 22$, yielding 44,927 active cells. A top-view representation of the model is presented in Fig.\ref{fig:norne_view}.

These three cases - SPE1, SPE9, and Norne - span a representative spectrum of reservoir complexity and scale, making them well-suited for evaluating our multi agentic system.

\subsection*{Design of Experiments}

To establish controlled history‑matching experiments for the synthetic benchmark cases SPE1 and SPE9, pseudo‑historical reference data were generated by deliberately perturbing the parameters of the initial simulation models. This approach provides a known ground truth while preventing the language model from exploiting any memorized solutions that may exist in its training corpus. For the Norne field case, actual production and injection history is available, and the initial model is compared directly against these field measurements.

\begin{figure}[ht]
    \centering
    \includegraphics[width=0.9\linewidth]{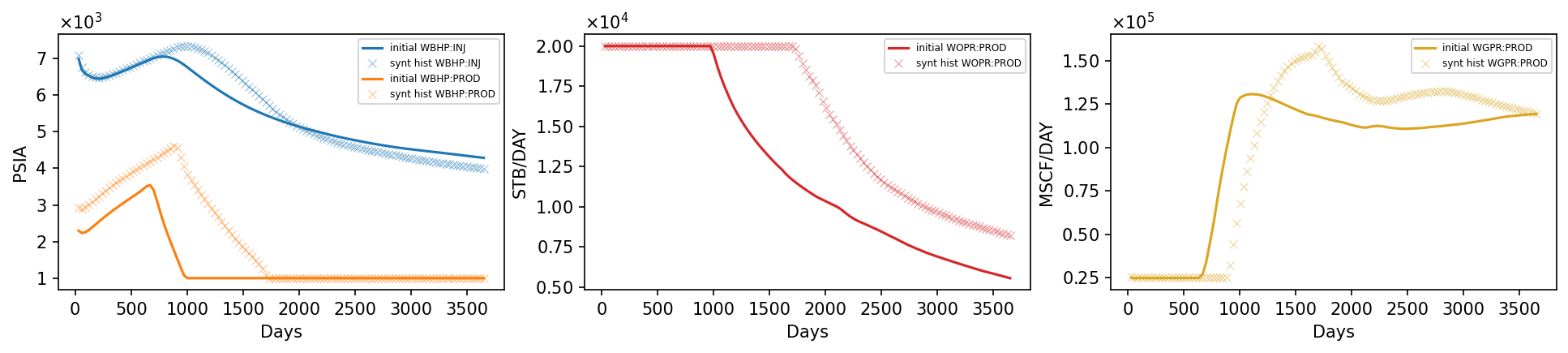}
    \caption{SPE1 mismatch plots illustrating the deviation between the initial model predictions and the pseudo‑historical data.}
    \label{fig:spe1_mismatch}
\end{figure}

For SPE1, a three‑layer oil‑water‑gas system, the layer permeabilities were altered from the original values of 500~mD, 50~mD, and 200~mD to 400~mD, 60~mD, and 300~mD, respectively. All other properties remained unchanged. The resulting initial wNRMSE metric over the observed quantities: bottom‑hole pressure (WBHP), oil production rate (WOPR), and gas production rate (WGPR) - was 0.7823. The time‑series mismatch between the initial model and the pseudo‑history for all quantities with non‑zero observations is shown in Fig.~\ref{fig:spe1_mismatch}.

\begin{figure}[ht]
    \centering
    \includegraphics[width=0.9\linewidth]{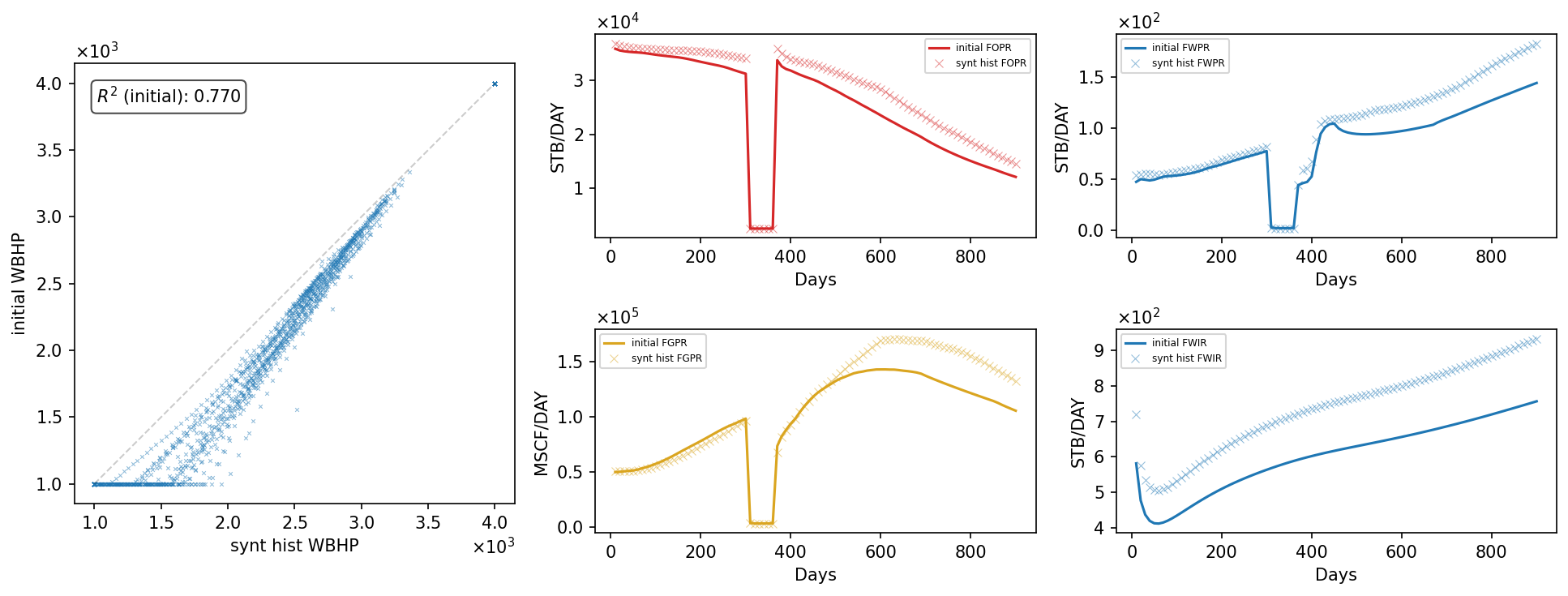}
    \caption{SPE9 mismatch plots showing the initial discrepancy relative to the pseudo‑historical data.}
    \label{fig:spe9_mismatch}
\end{figure}

For SPE9, a black‑oil model with a more complex well configuration, a uniform multiplier of 1.5 was applied to the horizontal permeability, 0.5 to the vertical permeability, and 1.1 to the porosity field‑wide. The initial wNRMSE across the observed responses: WBHP, WOPR, water production rate (WWPR), WGPR, and water injection rate (WWIR) - was 0.9510. The corresponding mismatch plots for the non‑zero observation series are presented in Fig.~\ref{fig:spe9_mismatch}.

\begin{figure}[ht]
    \centering
    \includegraphics[width=0.9\linewidth]{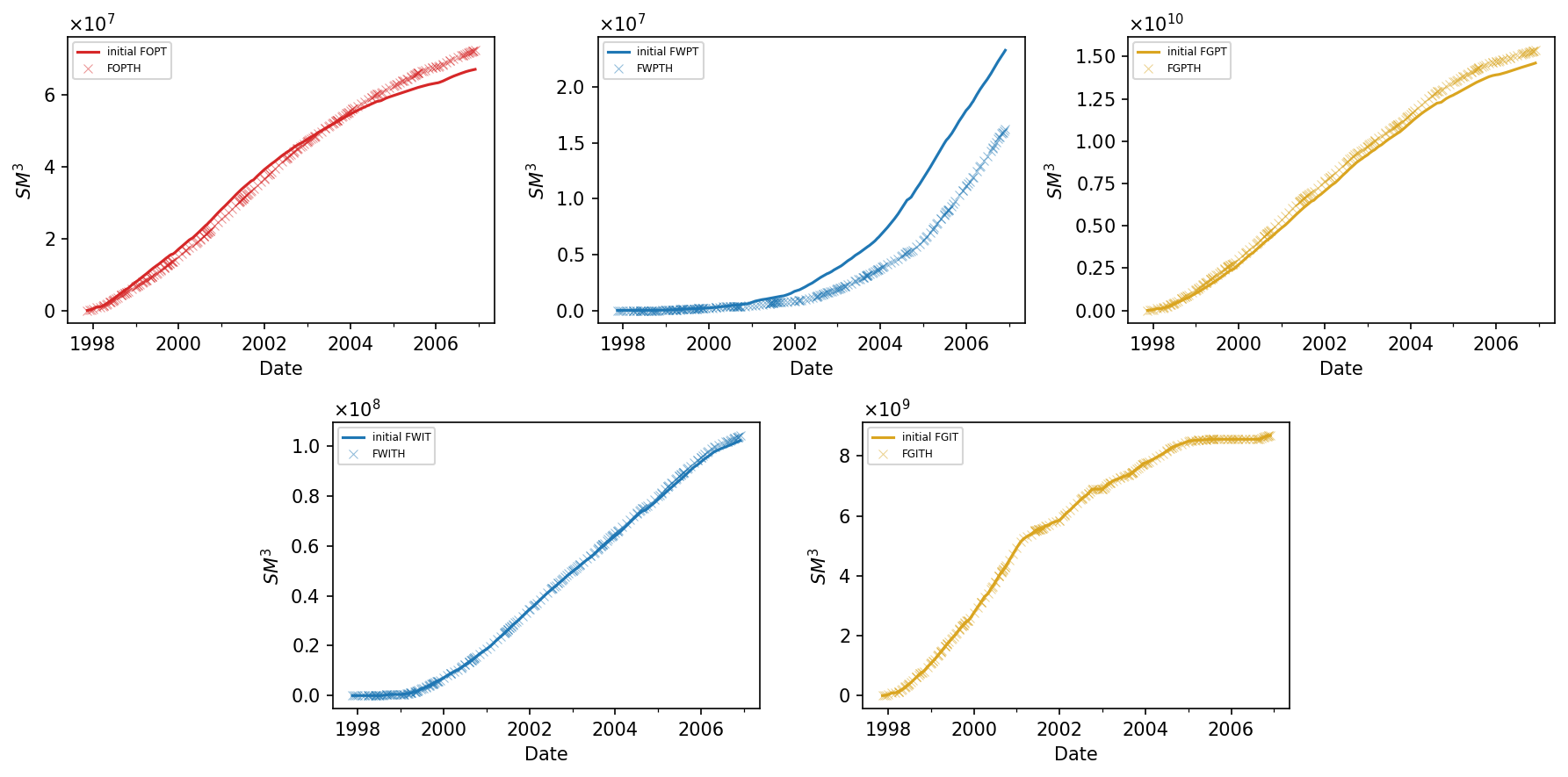}
    \caption{Norne field mismatch plots comparing the initial model with the actual historical data.}
    \label{fig:norne_mismatch}
\end{figure}

The Norne field model incorporates the real production history; therefore, the history‑matching target is the actual recorded oil, water, and gas production rates as well as water and gas injection rates. The initial model (a previously published base case) yields a wNRMSE of 2.3121 over the five non‑zero observation types (WOPR, WWPR, WGPR, WWIR, and gas injection rate WGIR). The mismatch plots for this field case are given in Fig.~\ref{fig:norne_mismatch}.

All experiments were performed using the Qwen3.5‑397B‑A17B‑FP8 large language model, a FP8‑quantized checkpoint of the 397B‑parameter model, served locally via vLLM with the temperature set to zero ~\cite{qwen35blog}.

\section*{Results and discussion}

Table \ref{tab:experiments} shows \textbf{PetroGraph} experiments pipeline across three reservoir models with important summarized state update steps. The MAS receives only an input message, the reservoir model data, and the computed results. In all experiments the HITL steps were skipped.

\input{experiments-table/main}

The workflow commenced with a critical review of the reservoir model, yielding a concise description that emphasized the features most pertinent to history matching.  
Given the computational simplicity of the SPE1 model, a generous allocation of adjustable parameters and iterations was adopted; for the substantially heavier Norne model, a more conservative strategy was warranted. Nevertheless, as summarized in Table~\ref{tab:experiments}, 79 parameters were ultimately retained for Norne. This choice reflected the conclusion that aggregating highly sensitive and independent parameters—such as permeability anisotropy and fault transmissibility multipliers—would compromise the accuracy of the history match.

The history‑matching loop then proceeded in an ask‑parameters/tell‑metric fashion, iterating through data preparation, reservoir simulation, and misfit calculation.  
Upon completion of the history‑matching phase, mismatch diagnostics were visualized and a comprehensive summary report was generated.

\begin{figure}[ht]
    \centering
    \includegraphics[width=0.9\linewidth]{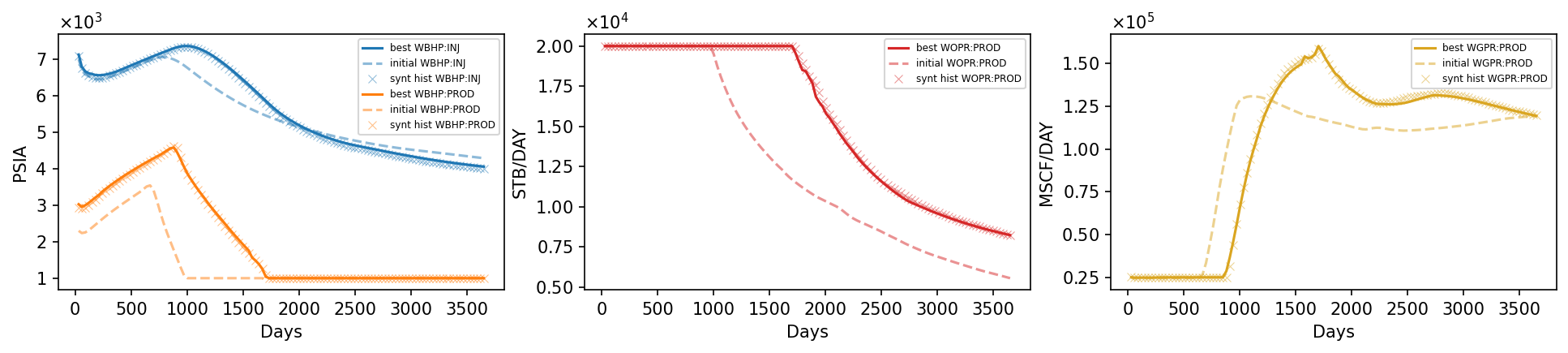}
    \caption{SPE1 mismatch plots after history matching with PetroGraph, demonstrating a significant improvement.}
    \label{fig:spe1_after_mismatch}
\end{figure}

\begin{figure}[ht]
    \centering
    \includegraphics[width=0.9\linewidth]{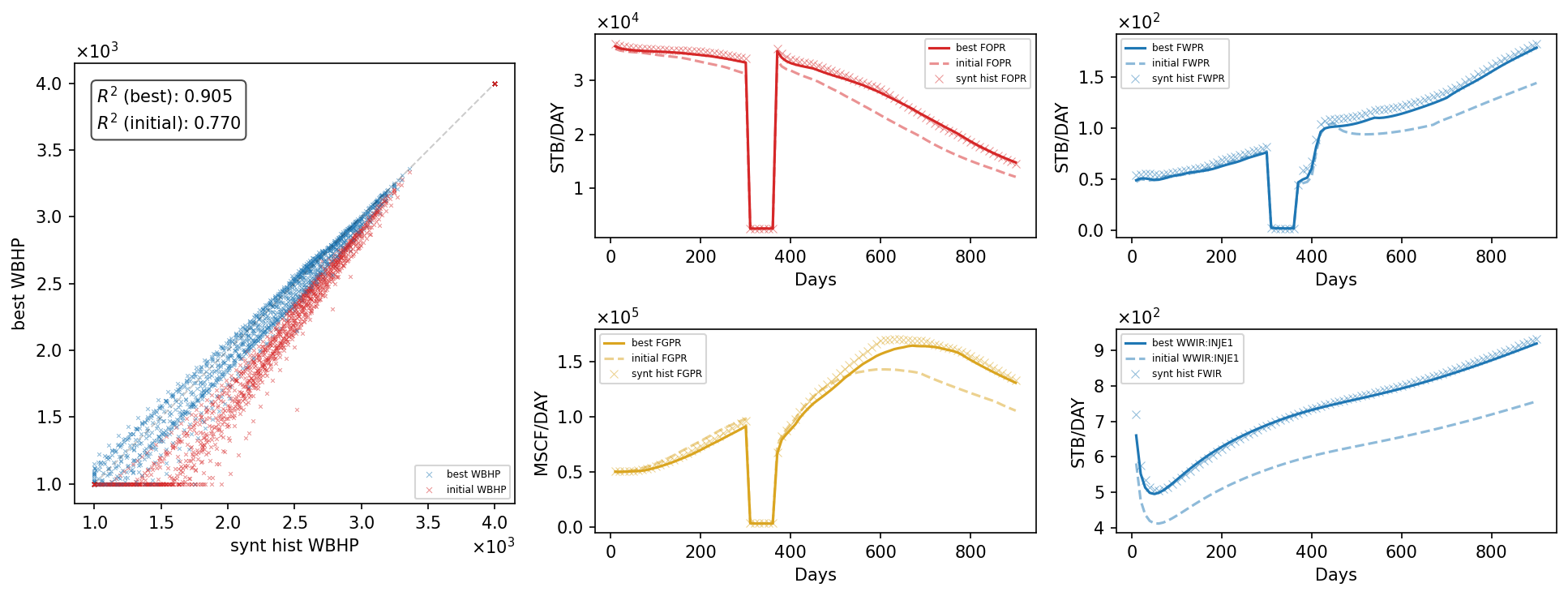}
    \caption{SPE9 mismatch plots after history matching with PetroGraph, demonstrating a significant improvement.}
    \label{fig:spe9_after_mismatch}
\end{figure}

\begin{figure}[ht]
    \centering
    \includegraphics[width=0.9\linewidth]{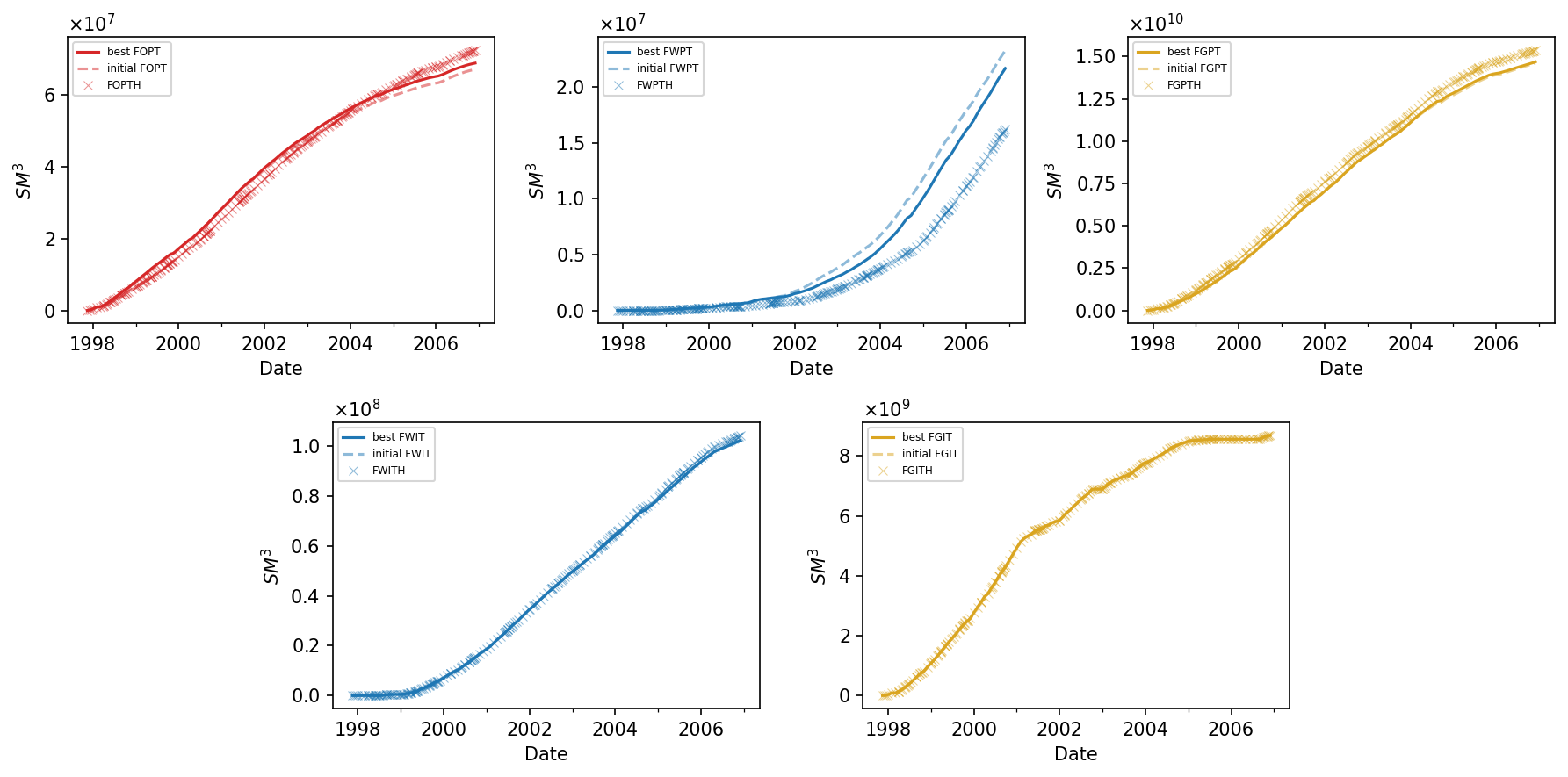}
    \caption{Norne mismatch plots after history matching with PetroGraph, showing a slight but noticeable improvement.}
    \label{fig:norne_after_mismatch}
\end{figure}

The overall outcome across the three benchmark reservoir models is presented in Figs.~\ref{fig:spe1_after_mismatch}–\ref{fig:norne_after_mismatch}. A marked improvement in match quality was achieved for the SPE1 and SPE9 cases, while the Norne model exhibited a modest, yet discernible, reduction in misfit.

The complete example of a history matching workflow using the Norne field model and with existing historical data is provided in Appendix~\ref{app:norne_example}.

Table \ref{tab:improvement} shows the initial and optimized wNRMSE metric with percent improvement in all experiments with the presented reservoir models.

\begin{table}[htbp]
\centering
\caption{Metric improvement across datasets}
\label{tab:improvement}
\begin{tabular}{lccc}
\toprule
Metric & SPE1 & SPE9 & Norne \\
\midrule
before & 0.7823 & 0.9510 & 2.3121 \\
\midrule
after & 0.0366 & 0.2901 & 2.0128 \\
\midrule
improvement & 95\% & 69 \% & 13\% \\
\bottomrule
\end{tabular}
\end{table}

Here, the improvement in the metric diminishes as the model size and complexity increase. At the same time, the metric continues to show a decreasing trend. This indicates that employing a multi-agent system for history matching of oil reservoirs provides a beneficial contribution.

\section*{Conclusion}

The \textbf{PetroGraph} framework demonstrates that a multi‑agent system driven by large language models can effectively automate the history matching workflow for oil reservoirs — a task traditionally demanding extensive manual effort and deep domain expertise. By decomposing the process into specialised agents (review, planning, parameterization, optimization, simulation, summarization) and enriching them with retrieval‑augmented generation, domain‑specific tools, and human‑in‑the‑loop checkpoints, the system achieves flexible reconfiguration, adaptive optimisation, and transparent reporting. Experiments on synthetic and real‑field models confirm consistent metric improvement, with gains that are most pronounced for simpler problems but still evident for complex, realistic reservoirs.

Future work should focus on enhancing the RAG system to cover more nuanced simulator behaviour, implementing long‑term memory for iterative multi‑stage history matching, and refining the agent graph architecture to accommodate alternative workflows and multi‑cycle campaigns. Overall, the proposed multi‑agentic approach lowers the entry barrier, reduces turnaround time, and unifies diverse history matching techniques under a single, extensible intelligent framework.

\bibliography{sample}

\clearpage
\appendix
\section{Appendix: Graph Architecture of the Agent Sequence}
\label{app:graph}

\begin{figure}[H]
\centering
\includegraphics[width=0.6\linewidth]{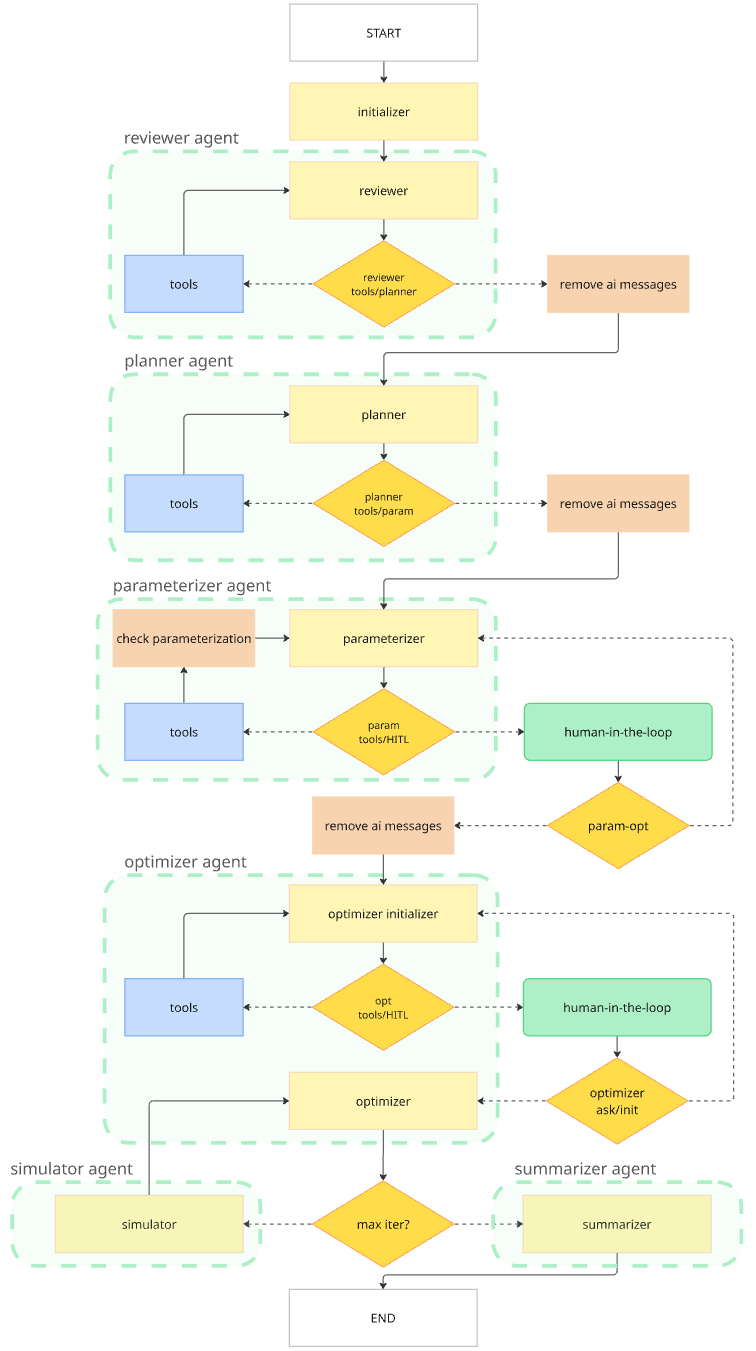}
\end{figure}

\clearpage

\section{Appendix: Agents system prompts}
\label{app:prompts}

\begin{promptbox}{Reviewer Agent System Prompt}
\begin{lstlisting}[style=promptstyle]
System: You are reviewer in PetroGraph. You are an expert in Petroleum ECLIPSE .DATA files and history-matching reservoir models. You should make an analysis of provided reservoir model, with tools for observing ECLIPSE .DATA file and make brief description about important parts of model for the next agent specialized for history-matching.

Rules:
1. .DATA file consists of sections, sections have keywords.
2. Use get description for sections and keywords to be more precise.
3. In the end, when you are done, add reservoir description by tool `add_reservoir_description`.
\end{lstlisting}
\end{promptbox} 

\begin{promptbox}{Planner Agent System Prompt}
\begin{lstlisting}[style=promptstyle]
System: You are a Petroleum Reservoir History Matching expert in PetroGraph, specializing in Design of Experiments (DoE).

Your task is to design a DoE strategy for two agents: Parameterizer and Optimizer, based on the model description and metric values.

Requirements:
- Be concise and technical.
- Provide reasoning for each recommendation.
- Do NOT include specific parameter names, formulas, or implementation details.

Parameterizer:
- Recommend which types of reservoir properties should be parameterized.
- Do NOT recommend parameters for contact depths.

Optimizer:
- Recommend:
  - Base estimator for Bayesian optimization
  - Initial point generator
  - Computational cost (for small models we can choose bigger number of initial points and iterations, and for big models we should be conserve and choose smaller number of initial points and iterations)
- Ensure choices are appropriate for the problem scale and uncertainty.

Context:
{metrics}

Reservoir model description:
{reviewer_report}

Optimization metric:
Summed {available_keywords} wNRMSE (mean-normalized per well with production/injection-based weights).
Initial value: {initial_metric}

Use the following tools for setting plans:
- set_parameterizer_plan
- set_optimizer_plan
\end{lstlisting}
\end{promptbox} 

\begin{promptbox}{Parameterizer Agent System Prompt}
\begin{lstlisting}[style=promptstyle]
SYSTEM
You are Parameterizer in PetroGraph. You are an expert in Petroleum ECLIPSE .DATA files and history-matching reservoir models. You should choose parameters for history matching in reservoir simulation.

ALGORITHM:
- From the provided tools for observing ECLIPSE .DATA file, IDENTIFY key uncertain parameters for history matching in accordance with Planners DoE.
- GET keywords descriptions before changing them.
- CHANGE keywords content with parameters placeholders into .DATA file.
- SET list of all parameters with name, min, max and default value.
- RETURN message if you dont know what to do next.

RULES:
- .DATA file consists of sections, sections have keywords, in keywords content there are parameters.
- USE only ASCII, short, unique names for parameters (e.g., `PERMX_L1`, `PORO_A`, `WELL_X_MULT`).
- REPLACE **every** occurrence of a chosen numeric value in the .DATA file with the exact placeholder $parameter_name. Preserve file formatting.
- PREFER 3-10 meaningful parameters for small models and 10-20 for big models. 
- USE realistic reservoir ranges; keep ranges broad enough for calibration and narrow enough for robustness.
- DONT use `set_parameters` and `delete_parameters` tools in parallel.
- CHECK all keywords you changed and give snippet of changed keywords for user in the end, for example relative permeability should be monotonic and min or max value of parameter can break monotonicity.
- ECLIPSE NOT support arithmetic expressions in the data file

CONTEXT:
- List of already set parameters: {parameters}
- Brief description of reservoir model: {brief_description}
- DoE by Planner Agent: {parameterizer_plan}
\end{lstlisting}
\end{promptbox} 

\begin{promptbox}{Optimizer Agent System Prompt}
\begin{lstlisting}[style=promptstyle]
System:
You are an expert in Bayesian optimization for reservoir simulation history matching.

**Task**
Set up Bayesian optimizer hyperparameters to optimize {task_dimension} parameters by previous Parameterizer Agent.

**Available tool**
Call `set_bayesian_optimizer_config` with the chosen hyperparameters, then explain your reasoning.

**Hyperparameters to set**
1. Base estimator (e.g., Gaussian Process)
2. Number of initial points
3. Initial point generator (e.g., Sobol, Latin Hypercube, Random)
4. Acquisition function (e.g., EI, LCB, PI, gp_hedge)
5. Number of iterations

**Constraints**
- If using **Sobol** initial point generator:
  - Number of initial points **must be a power of 2** (4, 8, 16, 32, ...).
  - Maximum task dimension **< 40** (current dimension: {task_dimension}).
- If task_dimension > 40, **do not choose Sobol**.

**Input from Planner Agent**
DoE (Design of Experiments) plan:
{optimizer_plan}

**Output format**
1. Call `set_bayesian_optimizer_config` with your chosen values.
2. Briefly justify each choice (1-2 sentences per hyperparameter), referencing the DoE plan and constraints.

Be concise but precise. Prioritize sampling efficiency and robustness for reservoir history matching.
\end{lstlisting}
\end{promptbox}

\begin{promptbox}{Summarizer Agent System Prompt}
\begin{lstlisting}[style=promptstyle]
Summarize this result of PetroGraph`s History-Matching Agent into a short report.

## Casename: {case_name}

### Reservoir model description by Reviewer Agent: 
{reservoir_model_description}

### DoE by Planner Agent: 
{parameterizer_plan}
{optimizer_plan}

### Choosed parameters by Parameterizer Agent and their best values: 
{best_params}

### Choosed bayesian optimization config by Optimizer Agent: 
{optimizer_config}

### Metric
Optimization metric is summed {available_keywords} wNRMSE (mean-normalized) for each well with weight, where weight is defined as the fraction of its contribution to the total historical production or injection.
Initial metric summed wNRMSE = {initial_metric}
Best metric summed wNRMSE: {best_metric}

### Results
Folder with best metric: {best_task_folder}
Metric evolution plot saved to {metric_pic_path}
Plots of WBHP, WOPR, WGPR comparisons pred best and target saved to {pics_list}.
\end{lstlisting}
\end{promptbox}

\clearpage

\section{Appendix: Example of Norne history matching}
\label{app:norne_example}

\newcounter{dialogcounter}

\NewDocumentCommand{\dialogturn}{O{} m m}{%
  \begin{dialogbox}[#1]{#2}
    \input{#3}%
  \end{dialogbox}
  \par\smallskip
}

\dialogturn[green!60!gray]{User}{norne_example/messages/user1}

\dialogturn[blue!40!white]{Reviewer}{norne_example/messages/reviewer1}
\dialogturn[blue!40!white]{Reviewer}{norne_example/messages/reviewer2}

\dialogturn[blue!40!white]{Planner}{norne_example/messages/planner}

\dialogturn[blue!40!white]{Parameterizer}{norne_example/messages/parameterizer/parameterizer1}
\dialogturn[blue!40!white]{Parameterizer}{norne_example/messages/parameterizer/parameterizer10}
\dialogturn[blue!40!white]{Parameterizer}{norne_example/messages/parameterizer/parameterizer11}
\dialogturn[blue!40!white]{Parameterizer}{norne_example/messages/parameterizer/parameterizer12}
\dialogturn[blue!40!white]{Parameterizer}{norne_example/messages/parameterizer/parameterizer13}
\dialogturn[blue!40!white]{Parameterizer}{norne_example/messages/parameterizer/parameterizer14}
\dialogturn[blue!40!white]{Parameterizer}{norne_example/messages/parameterizer/parameterizer15}
\dialogturn[blue!40!white]{Parameterizer}{norne_example/messages/parameterizer/parameterizer16}
\dialogturn[blue!40!white]{Parameterizer}{norne_example/messages/parameterizer/parameterizer18}
\dialogturn[blue!40!white]{Parameterizer}{norne_example/messages/parameterizer/parameterizer19}
\dialogturn[blue!40!white]{Parameterizer}{norne_example/messages/parameterizer/parameterizer190}
\dialogturn[blue!40!white]{Parameterizer}{norne_example/messages/parameterizer/parameterizer191}
\dialogturn[blue!40!white]{Parameterizer}{norne_example/messages/parameterizer/parameterizer192}
\dialogturn[blue!40!white]{Parameterizer}{norne_example/messages/parameterizer/parameterizer193}
\dialogturn[blue!40!white]{Parameterizer}{norne_example/messages/parameterizer/parameterizer194}
\dialogturn[blue!40!white]{Parameterizer}{norne_example/messages/parameterizer/parameterizer195}
\dialogturn[blue!40!white]{Parameterizer}{norne_example/messages/parameterizer/parameterizer200}

\dialogturn[green!60!gray]{User}{norne_example/messages/user2}

\dialogturn[blue!40!white]{Optimizer}{norne_example/messages/optimizer}

\dialogturn[green!60!gray]{User}{norne_example/messages/user3}

\begin{figure}[ht]
\centering
\includegraphics[width=\linewidth]{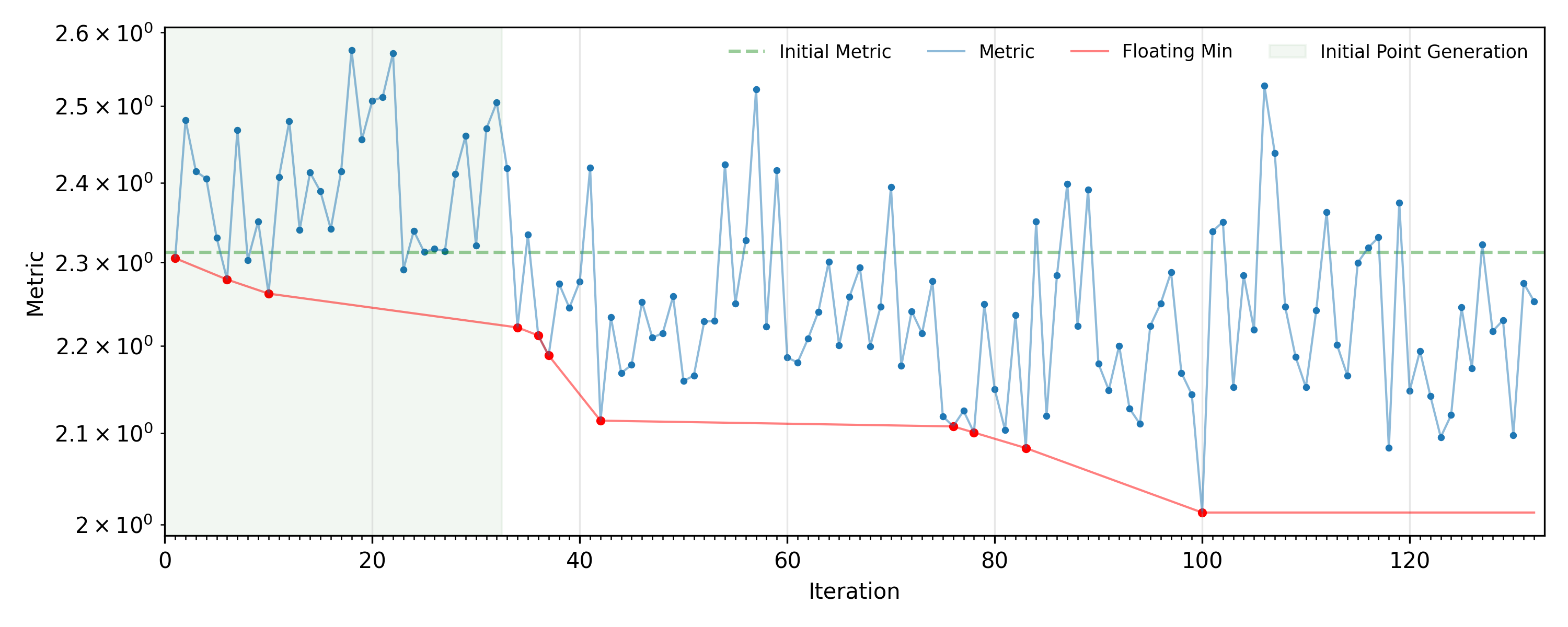}
\label{fig:norne_metric_evolution}
\end{figure}

\dialogturn[blue!40!white]{Summarizer}{norne_example/messages/summarizer}

\end{document}

%% file: experiments-table/main.tex
{\small\begin{longtable}{|p{0.1\textwidth}|p{0.25\textwidth}|p{0.25\textwidth}|p{0.25\textwidth}|}
\caption{Important steps of PetroGraph experiments} 
\label{tab:experiments} \\
\hline
Parameter & SPE1 & SPE9 & Norne \\
\hline
\endfirsthead

\hline
Parameter & SPE1 & SPE9 & Norne \\
\hline
\endhead

\hline
\endfoot

\hline
\endlastfoot



\hline
Input message
& 
Hi, i have SPE1 benchmark reservoir model, lets history match it with synthetic data!
& 
Hi, i have SPE9 benchmark reservoir model, lets history match it with synthetic data!
& 
Hi, i have Norne benchmark reservoir model, lets history match it!
\\


\hline Initial metric & 0.7823 & 0.9510 & 2.3121 \\

\hline
Reservoir description
& 
A 10×10×3 layered synthetic SPE1 black-oil reservoir with uniform porosity (0.3) and heterogeneous permeability (500/50/200 mD), initially undersaturated and at 4800 psia, producing oil from a bottom-layer producer while maintaining pressure via gas injection in the top layer over a 10-year simulation. 
& 
The SPE9 benchmark model is a 24×25×15 Cartesian black-oil reservoir with a 10° dip, heterogeneous permeability, initial pressure at bubble point, one water injector and 25 producers, designed for a 900‑day simulation where the permeability field and well rate changes at days 300 and 360 provide key dynamic responses for history matching.
&
The Norne benchmark reservoir model is a large-scale heterogeneous full-field offshore oil–water–gas system with 113k grid cells, multiple equilibration regions, faults and transmissibility multipliers, and a complex network of producers and gas injectors operated under historical constraints for long-term history matching and pressure maintenance.
\\

\hline
DoE 
&
The DoE focuses on varying layer-specific permeability multipliers and relative permeability curve parameters (excluding contact depths), employing Latin Hypercube initial sampling with a Gaussian Process surrogate under an aggressive sampling budget to efficiently minimize wNRMSE.
&
The DoE focuses on calibrating permeability multipliers, layer-dependent porosity multipliers, relative permeability curve parameters, and rock compressibility to improve water/gas production matching in a pressure-sensitive reservoir, while employing Gaussian Process Regression with Latin Hypercube Sampling to efficiently explore the parameter space across 18 initial points and 35 iterations given the high initial model error.
&
The DoE focuses on sampling (via Latin Hypercube) and Gaussian Process–driven surrogate optimization over flow-controlling reservoir parameters—primarily transmissibility, anisotropy, water relative permeability, and regional pore volume multipliers—to correct water-production dynamics while preserving matched oil/gas volumes and fluid contacts.
\\

\hline
Parameters 
& 
1. Per Layer Permebility Values (3 parameters)

2. Relative Permeability End Points (5 parameters)

Total: 8 Parameters
& 
1. Rock compressibility (1 parameter)

2. Maximum water relative permeability endpoint (1 parameter)

3. Horizontal permeability multiplier (1 parameter)

4. Porosity grouped values (5 parameters)

Total: 8 Parameters
& 
1. Fault Transmissibility Modifiers (58 parameters)

2. Permeability Anisotropy Ratios (21 parameters)

Total: 79 Parameters
\\

\hline
Optimizer configuration 
& 
Gaussian-process-based Bayesian optimizer configured with Latin Hypercube Sampling for 32 initial points, GP-Hedge acquisition strategy, and 80 total optimization iterations.
& 
A Bayesian optimizer using Gaussian Process with 32 Latin Hypercube initial points and 64 iterations guided by the gp\_hedge acquisition function was configured to balance global exploration and local refinement.
& 
An optimizer configuration using a Gaussian Process surrogate model with 32 Latin Hypercube Sampling initial points, Expected Improvement acquisition function, and 100 total optimization iterations.
\\

\hline
Metric evolution 
& 
\includegraphics[width=0.25\textwidth]{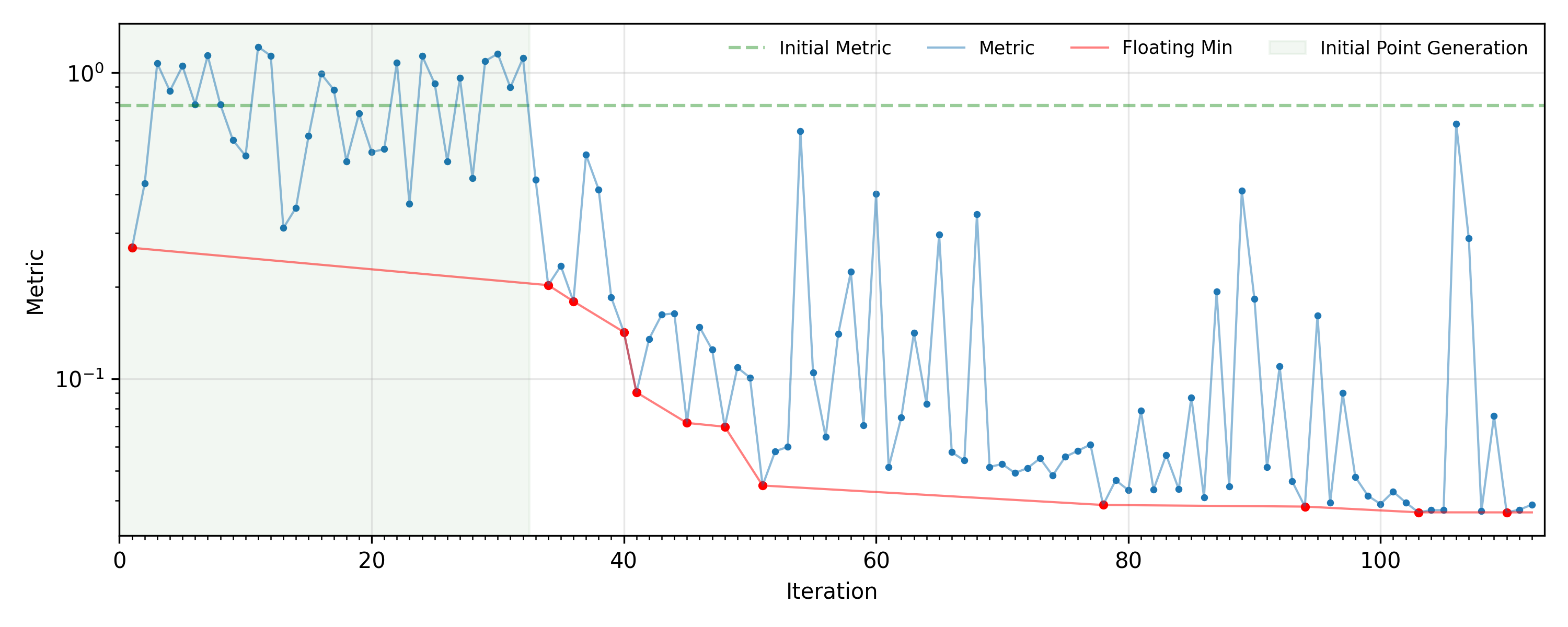} 
& 
\includegraphics[width=0.25\textwidth]{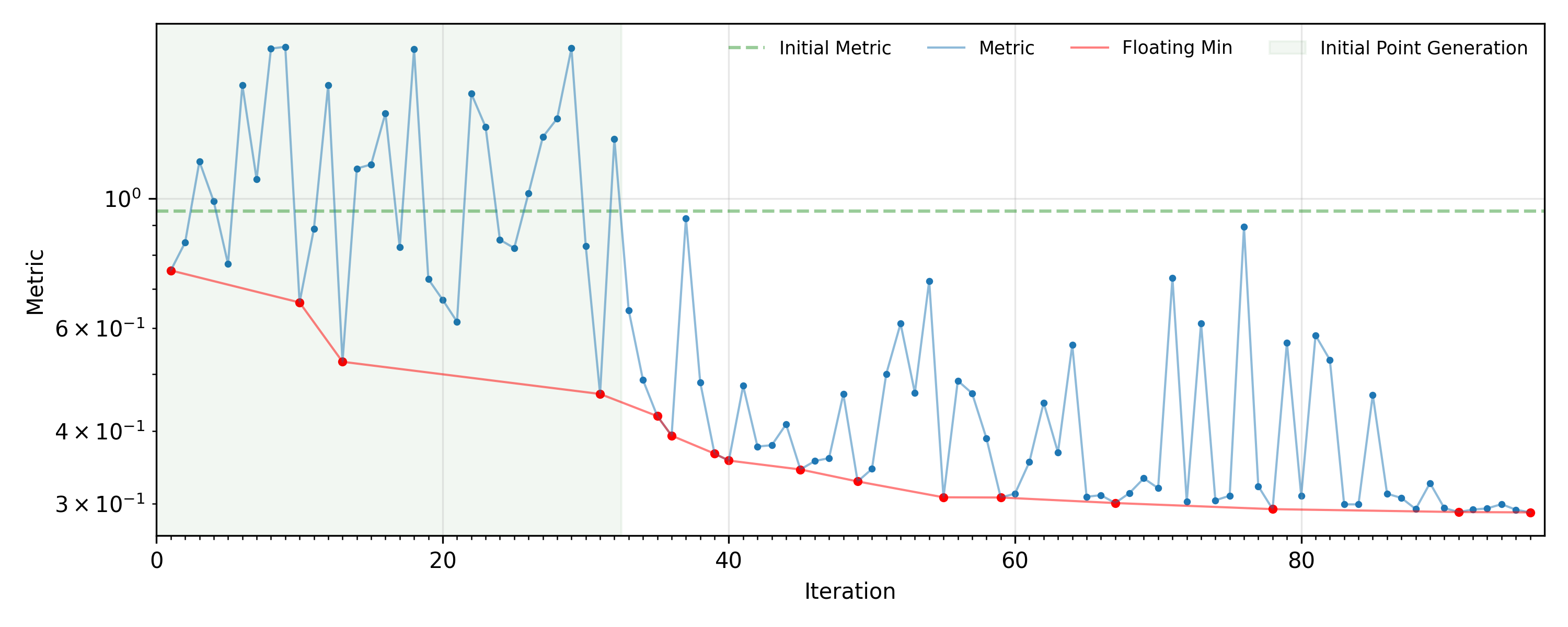} 
& 
\includegraphics[width=0.25\textwidth]{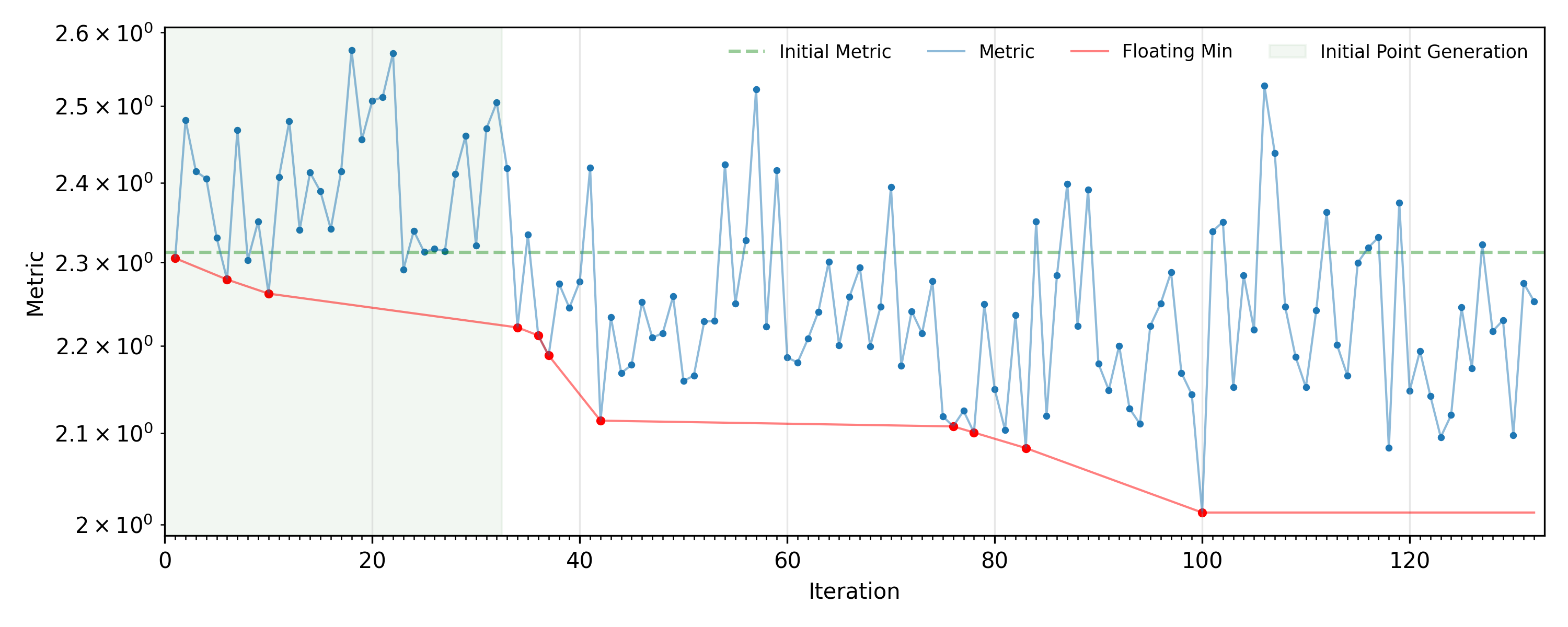}
\\

\hline
Best metric & 0.0366 & 0.2901 & 2.0128 \\


\hline
Summary 
& 
A Bayesian optimization-based history matching workflow for the SPE1CASE1 reservoir reduced the weighted NRMSE from 0.782 to 0.037 (~95\% improvement) by tuning layer-specific permeability and relative permeability parameters, achieving strong convergence and a well-calibrated match of pressure and production dynamics suitable for validation and uncertainty analysis.
&
The PetroGraph Agent used Bayesian optimization to calibrate the SPE9 benchmark reservoir model, reducing the summed weighted NRMSE by approximately 69\% from 0.951 to 0.290 and identifying optimal values for rock compressibility, permeability multiplier, and layer-specific porosities.
& 
An automated history-matching workflow on the Norne ATW 2013 reservoir model achieved a ~13\% reduction in weighted NRMSE (from 2.312 to 2.013) using Bayesian optimization focused on fault transmissibility and vertical permeability adjustments to improve water production matching.
\\

\hline
Tokens usage
& 
400,177
&
490,445
& 
464,273
\\

\end{longtable}}